# Invariant in variants

Cong Liu[1,2], Chen-Wu Wu[1*]

[1]Institute of Mechanics, Chinese Academy of Sciences, Beijing 100190, China.

[2]School of Engineering Science, University of Chinese Academy of Sciences, Beijing 100049, China.

*Corresponding author. Email: chenwuwu@imech.ac.cn & c.w.wu@outlook.com.

**Abstract:** The coronavirus Covid-19 mutates quickly in the pandemic, leaves people struggling to verify and improve the effectiveness of the vaccine based on biochemistry. Is there any physical invariant in the variants of such kind of pathogen that could be taken advantage to ease the tensions? To this point, extensive numerical experiments based on continuity mechanics were carried out to discover the vibration modes and the range of natural frequency of coronavirus Covid-19. Such invariant could help us in developing some flexible technique to deactivate the coronavirus, like as resonantly breaking the viral spike by ultrasound wave. The fundamental mechanisms governing such process are demonstrated via solving the coupled equations of acoustics and dynamics and thereafter the technique strategies proposed to efficiently realize the concept.

**Main Text:**

The war has been going on, between human and virus, focusing on the coronavirus Covid-19 these two years *(1-3)*. The tiny virus has been causing big trouble via changing into various mutants, of which at least nine variants (α, β, γ, δ, η, ι, κ, λ, μ) have been designated by WHO up to Aug 30, 2021 *(4, 5)*, all of which have been bringing so much sickness, injury and death to mankind. The human immune system could mostly conquer the invasive coronavirus autonomously with the aid of effective vaccine *(6)*, although which require continuing update of the vaccines to produce enough antibody *(7)*. Or else, we could actively deactivate the coronavirus in the ambient air *(8)* surrounding animals and human beings to largely suppress the infection probability. This could be realized by chemical way, like spraying strong acid/alkali disinfectant or physical approaches, such as electromagnetic wave/ ultrasonic wave irradiation *(9-12)*. Although, the overuse of chemical disinfectant is regarded to be partially responsible for accelerating the mutation of a large group of pathogens and increasing their antibiotic-resistances *(13)*. In contrast, it is almost impossible for a virus to mutate to escape from the attack by physical means, no matter through thermal or mechanical effects *(9, 10)* that could permanently devitalize the coronavirus via breaking its spike, membrane or even nucleic acid core in mechanical structure *(9, 14)*.

Despite the facts that the ultrasound wave could also be utilized to heat an object with thermally dissipating the mechanical energy transported in wave form *(14)*, ultrasound wave directly exciting the mechanical resonance of the virion would extremely deform and rupture its spike or envelope to quickly deactivate the virus *(16-18)*. Of course, such mechanical resonance of the viral structure with the stimulating ultrasound wave source rely fundamentally on the prerequisite that the frequency of the ultrasound wave is close enough to one of the natural frequencies of the viral structure *(17, 18)*, which is critical to avoid excessive energy loss due to the asynchronicity *(15)*. Therefore, accurate predictability of the natural frequency of the virion is figured on to develop any practical appliance that always operates in certain range of physical parameters such as frequency and power. This implies that the possible diversification in natural frequency of the virions would be a potential obstacle blocking the way of ultrasonic disinfection of coronavirus Covid-19, on which we firstly study the vibration modes of the viral structure with massive computations to involve the variations in both geometry patterns and mechanical properties of the viral mutants. Then, we demonstrate the stimulated stress field in the virion excited by the ultrasound wave of specific frequency, and finally proposed the strategies of frequency scanning to realize the concept.

The mechanical main body of the coronavirus Covid-19 in geometry commonly consists of an ellipsoidal/ spherical envelope shell and most probably 25~98 solid spikes nearly symmetrically distributed on the envelope *(18, 19)* as shown in Fig. 1, wherein the spherical envelope could also be approximated by a regular icosahedron *(19)*. Moreover, the axis length of the viral envelope is statistically revealed to be in the range of (60nm, 220nm), attached to which both the envelope thickness and the spike height are commonly in stable ratios to the average axis length *(19, 20)*. Such geometrical consistency of the virions truly demonstrate kind of mathematical similarity in the various mutants of the virus. Being mechanically made up of strong and tough protein, the viral envelope encloses and protects the carrier of genetic information, DNA/ RNA while the spikes work as intruder and deliverer of the DNA/ RNA into the target animal cells *(21-24)*.

Therefore, breaking the viral spike would prohibit the invasion into animal cell and hence the injection of the genetic information, which could be realized via imposing the ultrasound wave *(18)* rapidly if only the excitation frequency match well the natural frequency of the viral structure *(10, 18)*. From the point of view of physics, the deformation as well as stress would accumulate due to resonant excitation appears in a synchronized status and fractures arise once the stress approaches to the ultimate strength of the solid material *(25, 26)*. Accordingly, the key determinant is whether the natural frequency $\omega$ can be accurately evaluated, which is completed by solving the characteristic equations *(27)* related to the dynamic behaviors of the virion with the Finite Element Method *(28)*.

The typical vibration shapes along with natural frequency of the virus Covid-19, as shown in Fig. 2 have been calculated up to 120 and 500 orders for the viral configurations with 25 spikes and 98 spikes, respectively. Considering that fact that temperature fluctuation would change stiffness of the viral structure and therefore change the Eigen solutions to characteristic equation, the results on natural frequency have been maintained through the temperature range from 286.15 K to 329.15 K and form the continuous strips of data. It is revealed that the vibration modes of spike swing occupy up to about 50 orders of the vibration shapes of the virus with 25 spikes and up to about 196 orders of the virus with 98 spikes, respectively. The envelope vibration is negligible before the critical orders, which is the case regardless of the viral geometry of temperature level. It is indicated that the natural frequencies, $\omega_0^{(1)}/2\pi$ that associated with spike swing, basically fall into the ranges of (0.07GHz, 0.12GHz) or (0.26GHz, 0.44GHz), in which both ratios of upper bound to lower bound are less than 2. Such an invariant in natural frequency of the virions Covid-19 would provide a great maneuverability in frequency choosing to stimulate spike swing.

It is also reveled in Fig.2 that the viral envelop size determines the natural frequency of the virions and divides the frequency data into two separated regions, while as the effect of the envelop shape is relatively ignorable especially for the modes of spike swing. Without loss of generality, the dynamic responses of the spherical Covid-19 virions of diameter 220nm and 98 spikes to the planar ultrasound wave $P=P_0sin(t\omega_e)$ have been obtained via solving simultaneously the dynamic and acoustic equations again by Finite Element Method *(28)*.

Three strategies to excite the spike swing in the virions are demonstrated to apply ultrasound wave with $\omega_e$ being fixed, continuously-increased or stepwisely-increased, for which the wave profiles along with acoustic pressure and stress fields are plotted in the top of Figs. 3 and 4. Figures. 3 and 4 mainly present the stress magnifications $\sigma_m/P_0$ versus time with $\sigma_m$ being the maximum von-Mises equivalent stress *(26)* around the spike root, which would break once $\sigma_m$ equals to its' ultimate strength according to fracture theory *(26)*.

Results on each case in Figs. 3 and 4 have been separately plotted for the two temperature levels, 286.15K (in bright color) and 329.15K (in dark color) in different data strip whose thickness, the height difference in the values of upper bound and lower bound reflects thermal dissipation of mechanical energy due to damping. Of course, the damping would suppress the stress magnification in a whole while lead to temperature elevation in virions. It is found that the peak value of stress magnification is inversely related to temperature as temperature elevation always result in softening the material, namely reduce elastic modulus of the material.

The four groups of curves in Figs. 3 (a), (b), (c) and (d) depict $\sigma_m/P_0 \sim$ t for the cases that the ratio of excitation frequency $\omega_e$ of ultrasonic waves to the first natural frequency of the virion, $\omega_e/\omega_0^{(1)}$ is 0.5, 1.0, 1.5 and 2.0 in sequence. It is obviously shown in Fig. 3 (b) that the stress magnifications would be accumulated cyclically when $\omega_e$ equals to $\omega_0^{(1)}$, which indicates an apparent resonance stimulation. In this case, the peak stress magnification is over 100 before t=120ns if damping ratio is 0 and still over 70 at the instant t=100ns even if damping ratio is 0.01. That means, the maximum equivalent stress, $70 \times P_0$ around the spike root would approach to its' ultimate strength, about 0.141MPa *(12, 18)* within 100 Nano seconds even if the pressure altitude $P_0$ of the ultrasound wave is only 2 kPa, which is truly very low *(27)*. Meanwhile, the stress magnification would not be accumulated if the frequency ratio $\omega_e/\omega_0^{(1)}$ deviate from 1.0 as shown in Figs. 3 (a), (c) and (d) as the input mechanical energy would be largely lost.

Encouragingly, we found that a large magnitude of stress magnification always arises within 180 ns if change the frequency ratio $\omega_e/\omega_0^{(1)}$ from 0.5 to 2.0 during excitation, no matter continuously as in Fig. 4 (a) or stepwisely as in Fig. 4 (b). Moreover, the excitation via continuously changing frequency leads to a relatively smooth change in the peak values of stress magnification, while step-wisely changing results in relatively higher peak values with more rapid fluctuations. The peak stress magnifications within 180 ns are over 70 by both strategies for the cases of temperature 286.15K, in comparison the peak stress magnifications are over 50 for temperature 329.15K. The complete statistics show that the peak stress magnifications around the roots of all of the 98 spikes under any condition exceed 10 within 180ns, which still only request an ultrasonic wave of intensity lower than 15 kPa to break the viral spike.

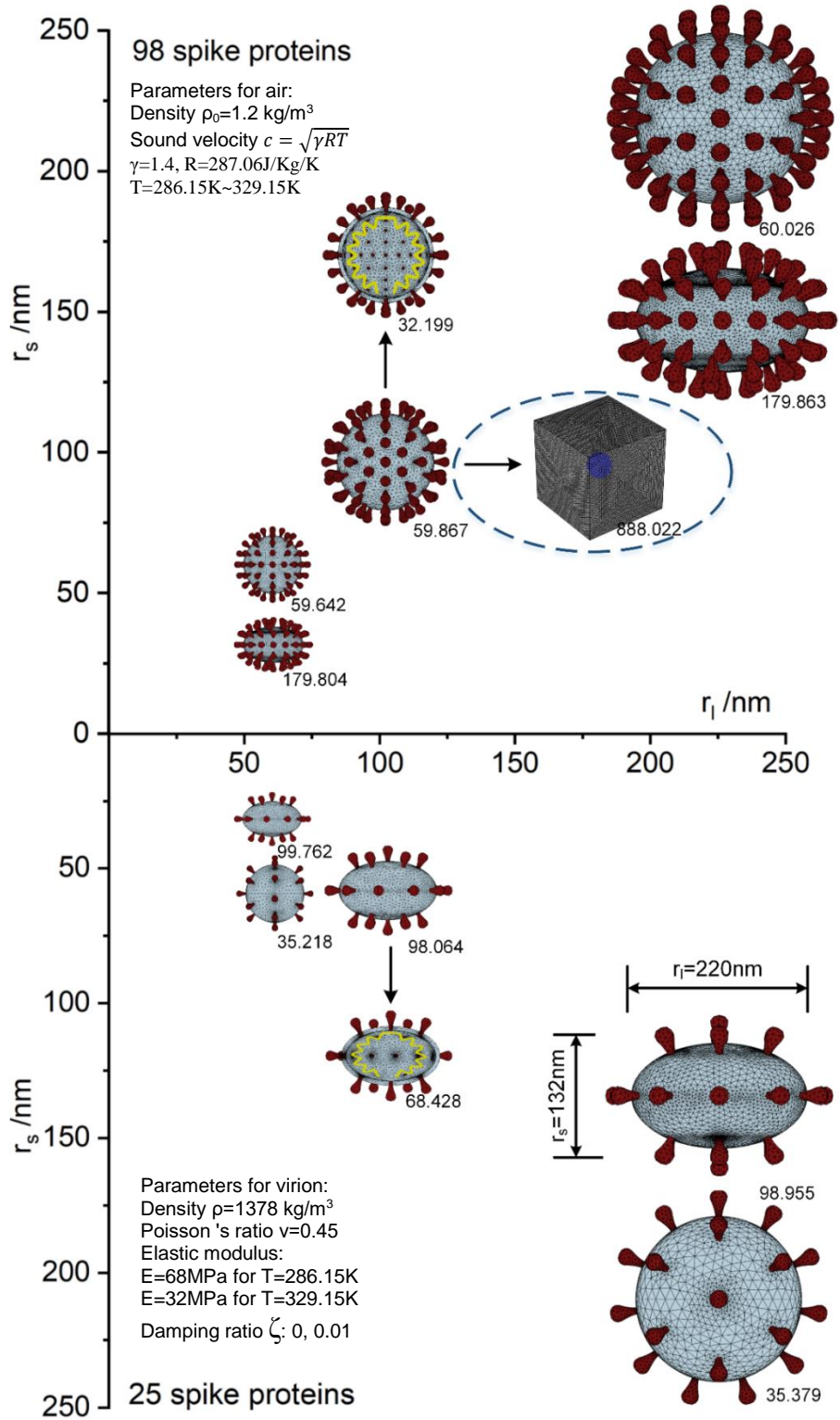

**Fig. 1 Sketch of the model on coronavirus Covid-19 with the lower part representing the virus with 25 spike proteins and upper part with 98 spike proteins, along with the parameters of the air and viral material used in computation (11, 12, 17, 18, 27).**

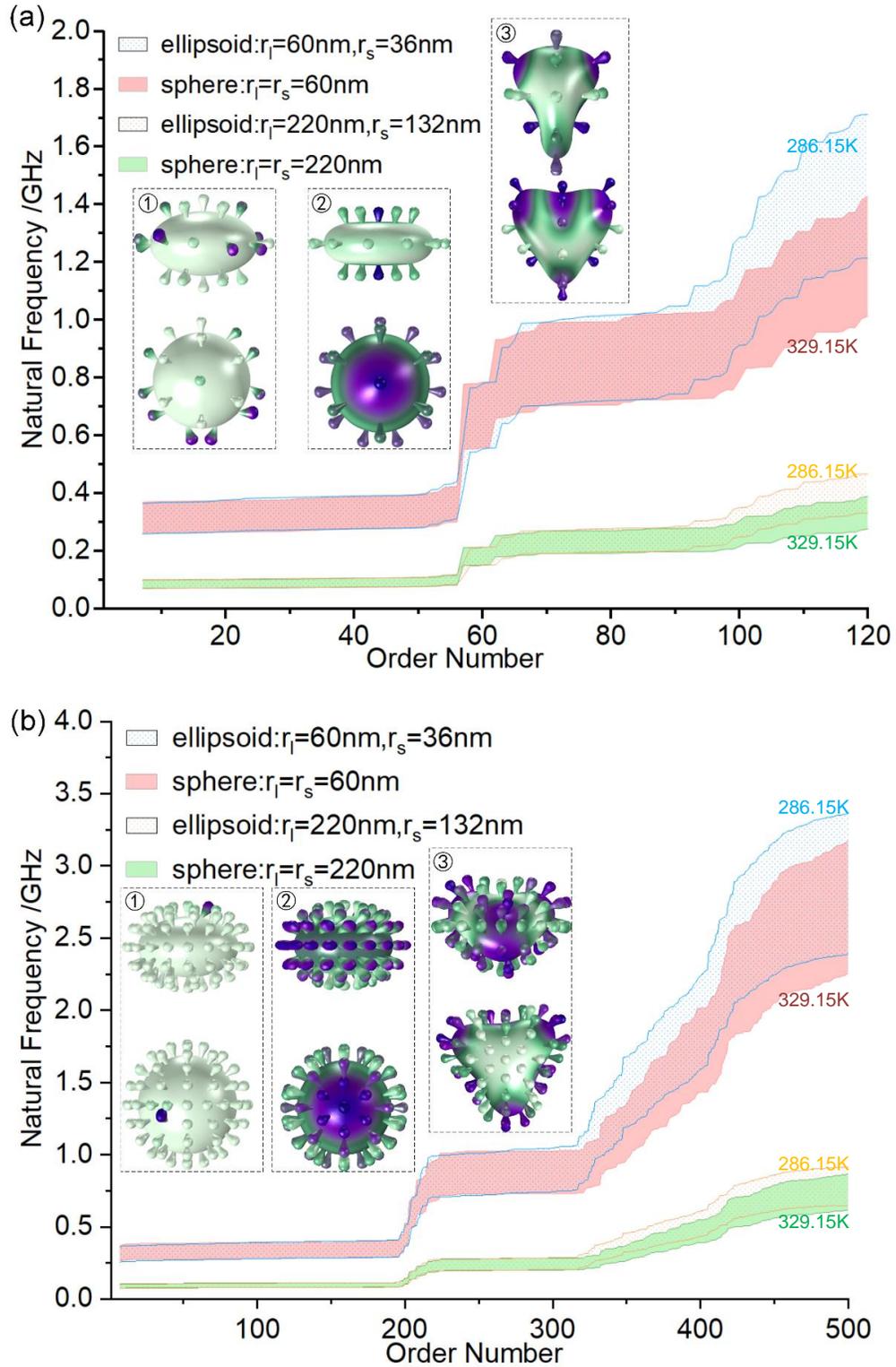

**Fig. 2. Natural frequencies of the virions of temperature in the range of (286.15K, 329.15K) with (a) the upper representing the virus with 25 spike proteins (up to 120th order) and (b) the lower with 98 spike proteins (up to 500th order), along with the first three distinctive vibration modes.**

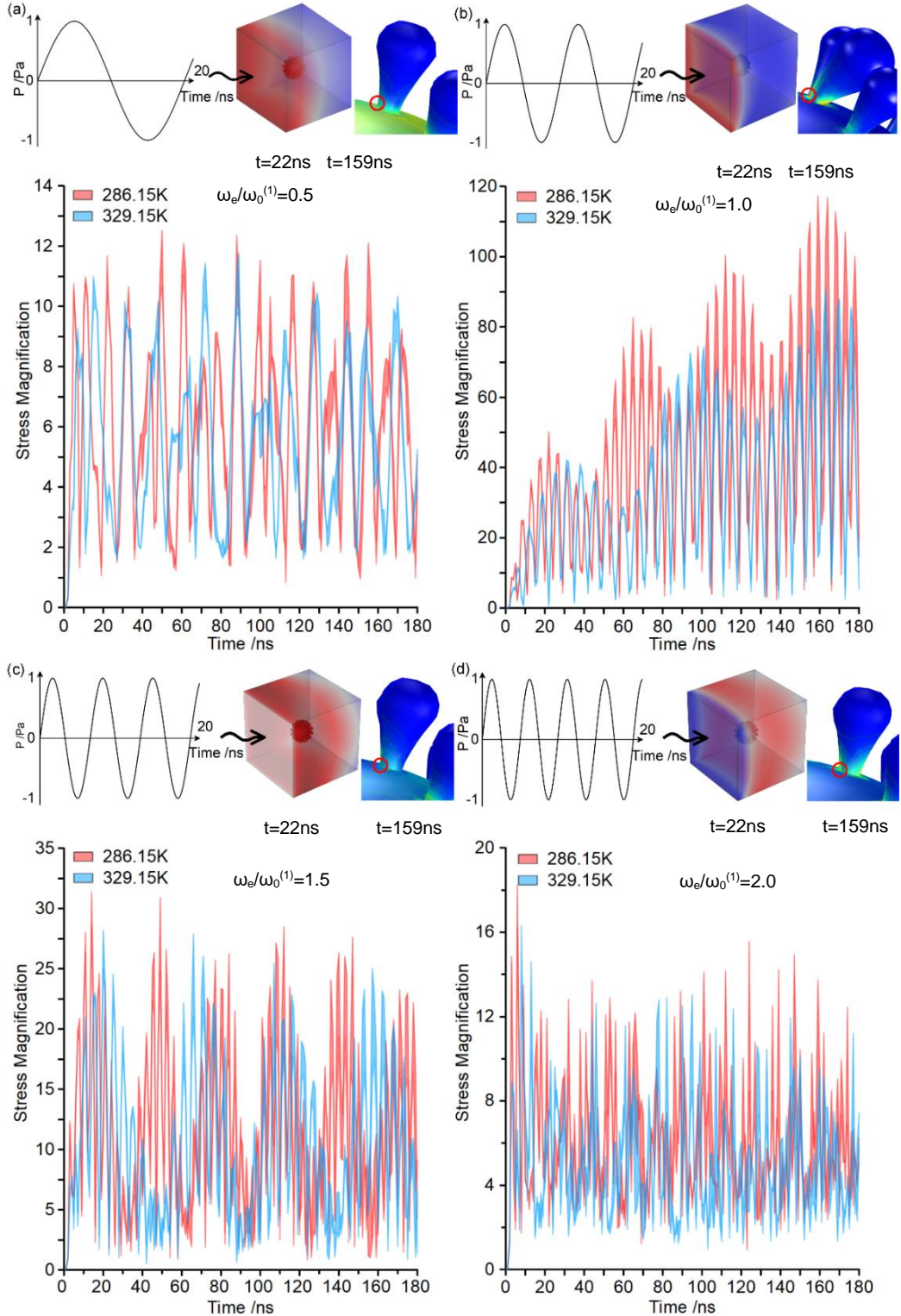

**Fig. 3.** Stress magnification around the root of the swung spike excited by ultrasound wave of single frequency $\omega_e$ that (a) 0.5, (b) 1.0, (c) 1.5 and (d) 2.0 times the first natural frequency of the virus $\omega_0^{(1)}$, along with acoustic pressure, stress field and the wave profiles of the ultrasonic waves for the cases of temperature T=286.15K.

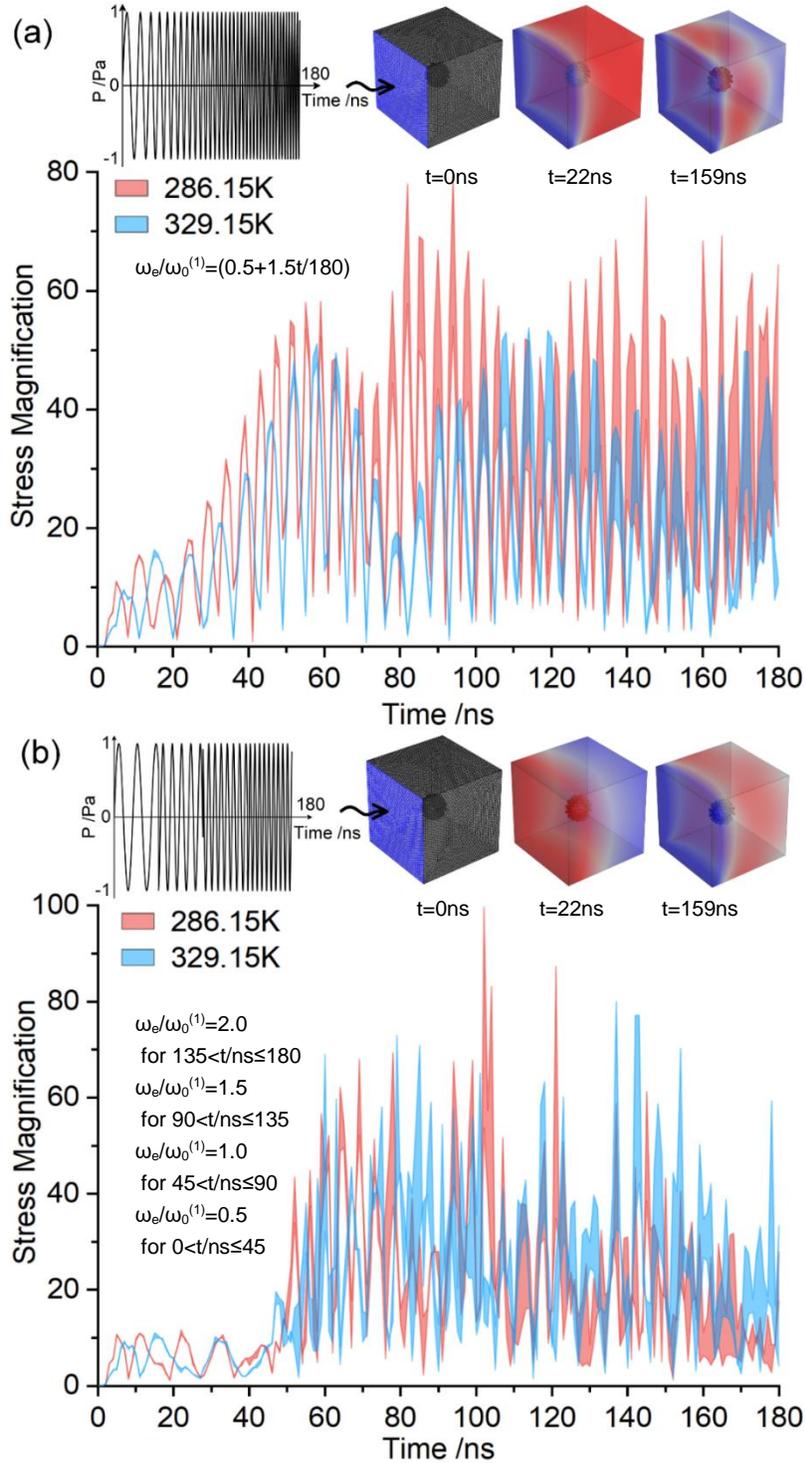

**Fig. 4.** Stress magnification around the root of swung spike excited by ultrasound wave of frequency $\omega_e$ that change (a) continuously and (b) step-wisely with time, along with acoustic pressure, stress field and the initial wave profiles of the ultrasonic waves for the cases of temperature T=286.15K.

**Author contributions:**

Conceptualization: CWW; Methodology: CWW, CL; Investigation: CL, CWW; Visualization: CL; Supervision: CWW; Writing – original draft: CWW, CL; Writing – review & editing: CWW

## More details of methods and more results

The investigation in this article mainly involves the materials of air and the virions, of which the physical parameters, such as density $\rho$, ratio of specific heat capacity $\gamma$, gas constant $R$, damping ratio $\zeta$, Poisson's ratio $\nu$ and elastic modulus $E$ for typical temperature levels if necessary have been cited from literature *(11, 12, 17, 18, 27)* and listed in Table S1.

The three-dimensional geometric models of the virions have been completed in CAD software through the three key procedures of shell feature generation, spike feature generation, and distributed assembly.

All of the ordinary/ partial differential equations have been numerically solved by Finite Element Method with the commercial program Comsol-Multiphysics, in which the cubic computation domain of size $1\mu m \times 1\mu m \times 1\mu m$ is accordingly discretized into nearly 1 million subdomains formed by the mesh grids as sketched in Fig. S1 (partitioned symmetrically).

Detailed mesh grid of the virions with 25 spikes and 98 spikes are shown in Figs. S2 and S3, in both of which the upper part and lower part sketch the mesh grid of the virion of spherical envelop and ellipsoidal envelop, respectively.

Natural frequencies and vibration modes are determined by the characteristic equations in structural dynamics (Eq. 1).

$$-\rho\omega^2\mathbf{u} = \nabla \cdot \boldsymbol{\sigma} \qquad \text{(Eq.1)}$$

Wherein, $\rho$ is the viral material, the operator $\nabla = [\frac{\partial}{\partial x}, \frac{\partial}{\partial y}, \frac{\partial}{\partial z}]^T$ with the superscript $T$ representing the transpose form of a vector, $\omega$ is the angular frequency, the eigenvectors $\mathbf{u}(x,y,z,t) = \mathbf{u}(x,y,z)e^{i\omega t}$ represents the displacement field for the case of time-harmonic response and (x, y, z) are the Cartesian coordinates.

The displacement field $\mathbf{u}$ determines the stress field $\boldsymbol{\sigma}$ through the geometrical equations (Eq. 2) governing strain-displacement relationship and constitutive equations (Eq. 3) governing stress-strain relationship *(25)*.

$$\varepsilon_{ij} = \frac{1}{2}\left(\frac{\partial u_i}{\partial x_j} + \frac{\partial u_j}{\partial x_i}\right); i, j = x, y, z \qquad \text{(Eq. 2)}$$

Wherein, $\varepsilon_{ij}$ are the components of strain tensor, $u_i$ are the displacement components.

$$\sigma_{ii} = \frac{E}{(1+\nu)(1-2\nu)}\left[(1-\nu)\varepsilon_{ii} + \nu(\varepsilon_{jj} + \varepsilon_{kk})\right]; \sigma_{ij} = \frac{E}{1+\nu}\varepsilon_{ij}; i, j, k = x, y, z \qquad \text{(Eq. 3)}$$

With $E$ being the elastic modulus of materials, $\nu$ the Poisson's ratio and (x, y, z) are the Cartesian coordinates.

The equations (Eq. 4) govern the movement, deformation and hence stress distribution of the virus particle, while equations (Eq. 5) control the propagation of the ultrasound wave in the media.

The boundary conditions (Eq. 6) are applied to the interface between the virion and the air to follow the law of conservation. The boundary condition (Eq. 7) is acted upon the side x=0 of the cubic volume of air with side length of $1\mu m$, at the center of which located the virus particle and the other sides of which is assumed as nonreflecting boundaries. The law of conservation and rule of continuity in energy and momentum are applied to the virus surface that acts as the interface between the virus particle and surrounding air environment.

$$\rho\frac{\partial^2\mathbf{u}}{\partial t^2} = \nabla \cdot \boldsymbol{\sigma} \qquad \text{(Eq. 4)}$$

$$\frac{1}{\rho c^2}\left(\frac{\partial^2 p}{\partial t^2}\right) + \nabla\left(-\frac{1}{\rho}\nabla p\right) = 0 \qquad \text{(Eq. 5)}$$

$$-\mathbf{n}\left(-\frac{1}{\rho c}(\nabla p)\right) = -\mathbf{n} \cdot \mathbf{u}_{tt} \qquad \text{(Eq. 6)}$$

$$p|_{x=0} = p_0 sin(\omega_e t) \qquad \text{(Eq. 7)}$$

Wherein, ρ is the density of virion or air, *t* is the time, *c* is the wave velocity in the air enclosing the virion, **u** is the displacement field and its' second partial derivative to time **u**$_{tt}$ means acceleration; n is the unit normal vectors, *p* is the acoustic pressure developed by the ultrasound wave, the acoustic pressure exerted on the domain boundary x=0 is assumed as $p$(t)=$P_0$sin($\omega_e$t) and $\omega_e$ is the angular frequency of the ultrasonic wave.

The amplitude $P_0$ of the acoustic pressure induced by the ultrasonic wave was set as unit magnitude "1" to normalize the numerical results considering the intrinsic linearity of the problem described by equations (2) and (4), which would be retained even the complex interaction between the incident ultrasound and the virus particle is involved herein. The propagation of the ultrasound, including reflection and refraction upon the surface of virus particle, could be accurately enough approximated with the above linear equations for the cases of specified temperature as the pressure is as moderate as discussed in the present study. This would lead to a constant sound wave velocity for any specified temperature, *c* independent of the pressure in the air *(25, 26)*.

The alteration ways of the frequency $\omega_e/\omega_0^{(1)}$ with time utilized in exciting the swing vibration of the viral spike are plotted in Fig. S4, in which the black solid line represents the case of continuously-increased $\omega_e/\omega_0^{(1)}$ while the red dashed line of stepwisely-increased $\omega_e/\omega_0^{(1)}$. It could also be expressed by a function of time, t as (Eq. 8) and (Eq. 9).

$$\omega_e/\omega_0^{(1)}=(0.5+1.5t/180) \tag{Eq. 8}$$

for the case of continuously change and

$$\omega_e/\omega_0^{(1)}=2.0 \text{ for } 135<t/ns\leq180; \omega_e/\omega_0^{(1)}=1.5 \text{ for } 90<t/ns\leq135;$$
$$\omega_e/\omega_0^{(1)}=1.0 \text{ for } 45<t/ns\leq90; \quad \omega_e/\omega_0^{(1)}=0.5 \text{ for } 0<t/ns\leq45. \tag{Eq. 9}$$

for the case of stepwisely chage.

The results on convergence verification on mesh size along with the element quality histogram are shown in Fig. S5, in which the calculated natural frequencies by the finite element models with five grades of element refinement are compared and the final mesh size is chosen as normal when the relative difference is lower than 1%.

The typical vibration modes of the spike swing in the virions are shown in Figs. S6-S9, which include the swing modes of single spike as well as of group of spikes.

The Stress fields in the virion at typical instants are presented in Figs. S10~S17 for the cases with different loading ways, temperature levels and damping ratio values. Note that the magnitudes of the stresses are maintained under the condition of the ultrasonic wave amplitude $P_0$=1 Pa.

The maximum stress magnifications, $\sigma_m/P_0$ arising around the roots all of the spikes have been counted and plotted versus time, up to 180 ns in Figs. S18-S25.

**Fig. S1.**

Sketch of mesh grid of half part of the computation domains of both the virus and the air wrapping it.

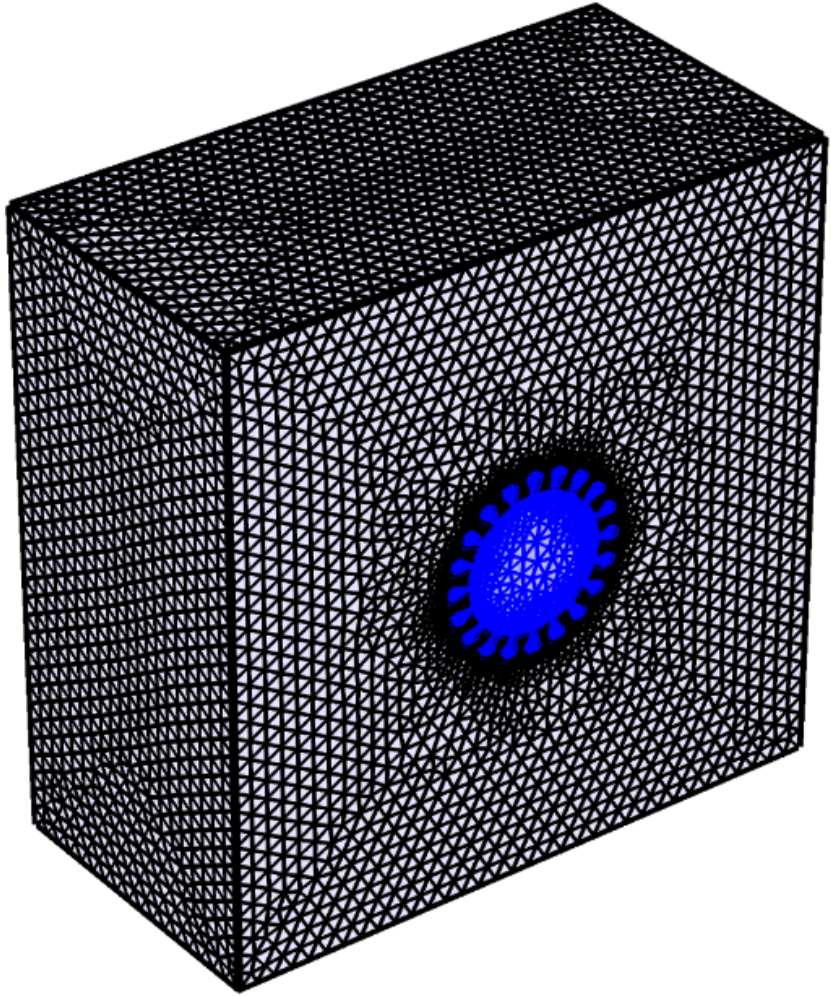

**Fig. S2.**
Sketch of mesh grid of the coronavirus Covid-19 with 25 spikes

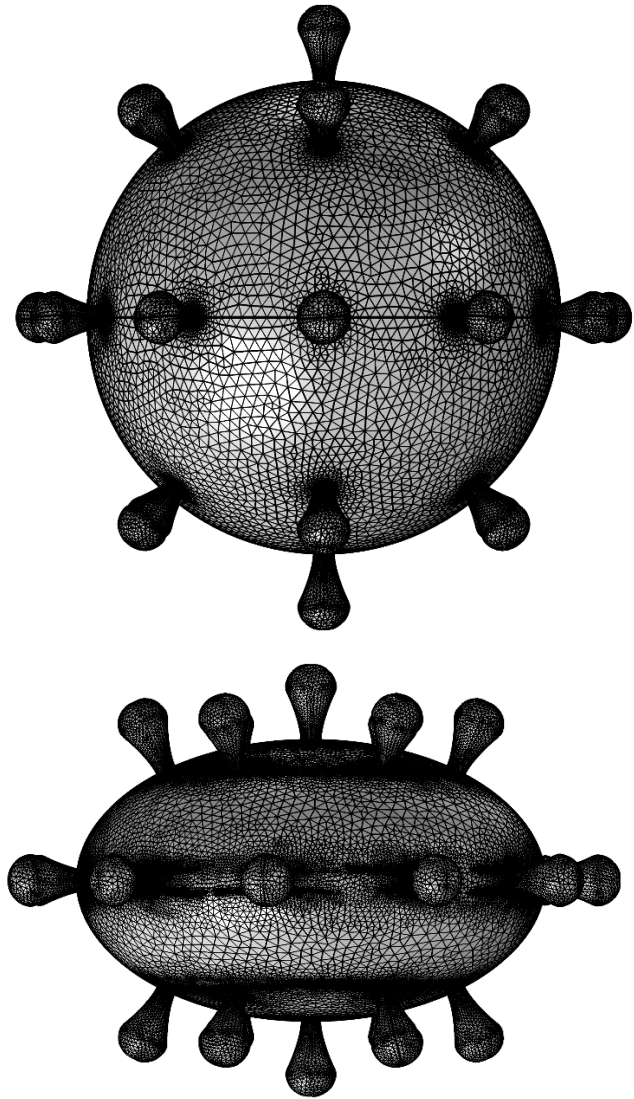

**Fig. S3.**

Sketch of mesh grid of the coronavirus Covid-19 with 98 spikes

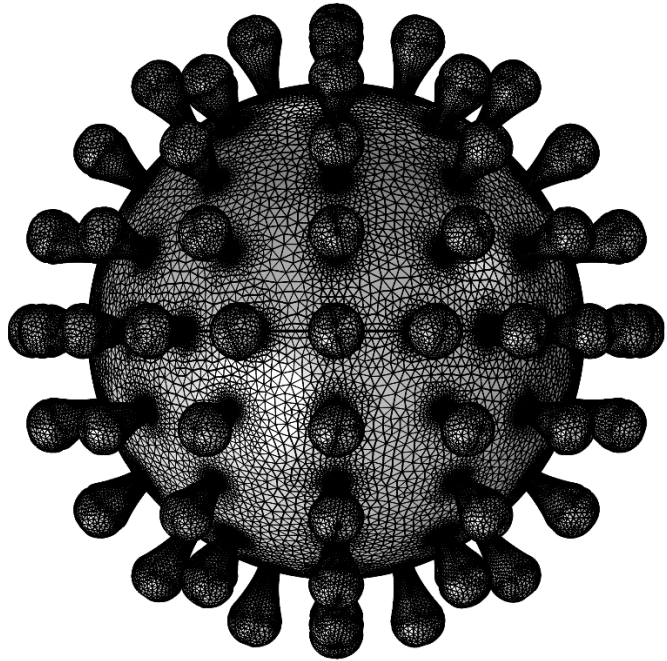

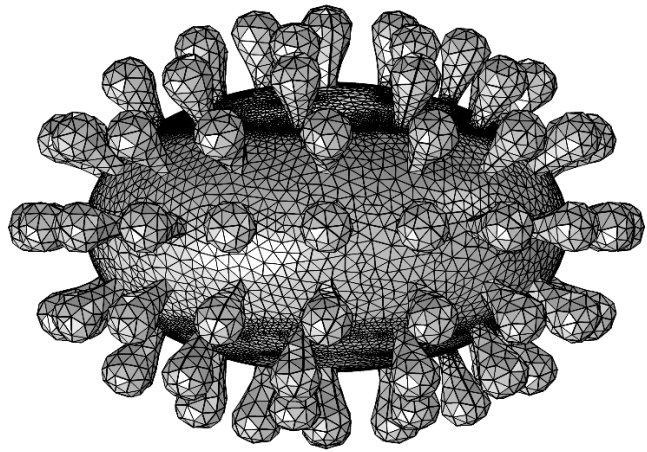

**Fig. S4.**

Sketch of the two kinds of ways changing frequency of the ultrasound wave

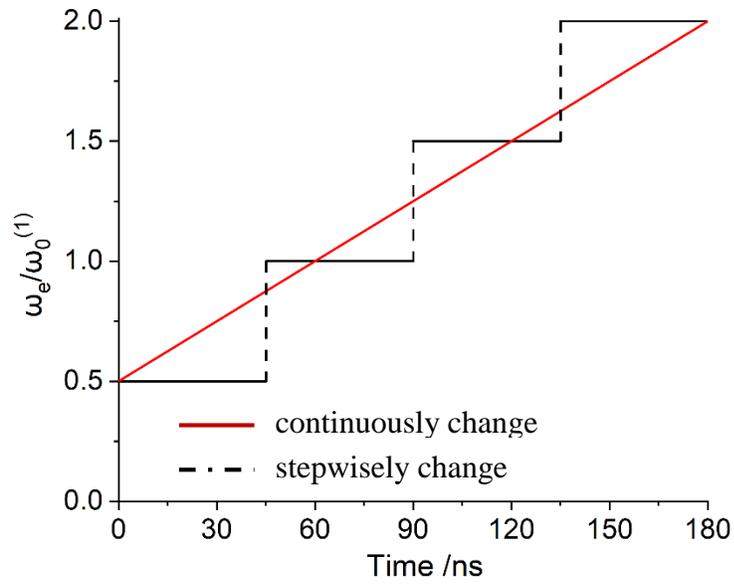

**Fig. S5.**

Outcome for the convergence verification on mesh size and the element quality histogram

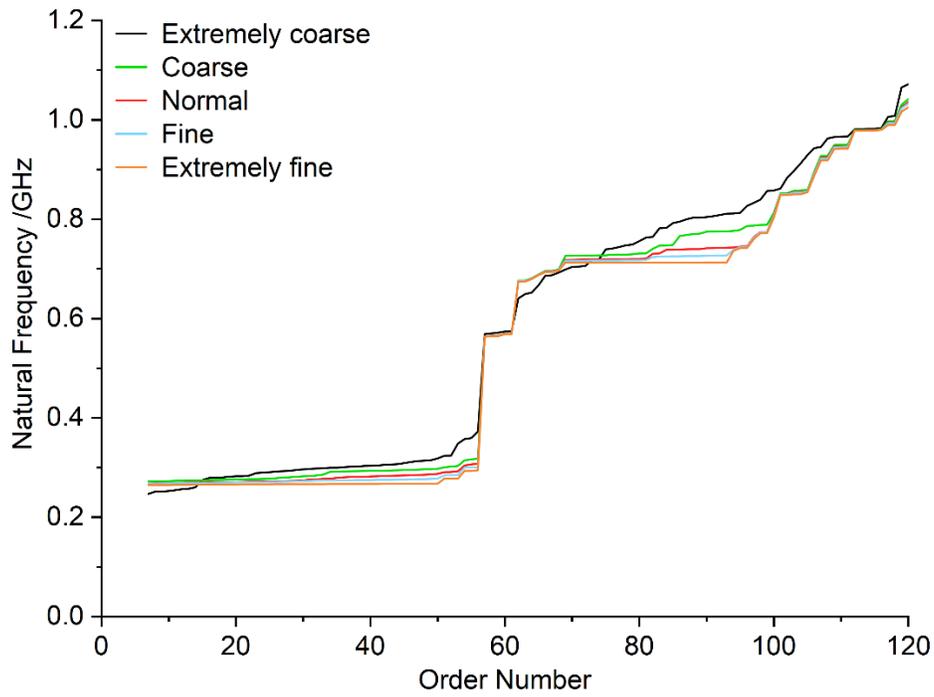

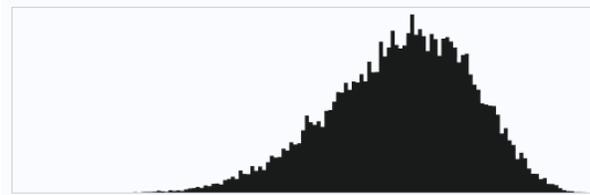

Element Quality Histogram for Normal Grade

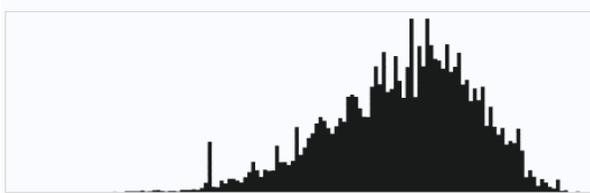

Element Quality Histogram for Coarse Grade

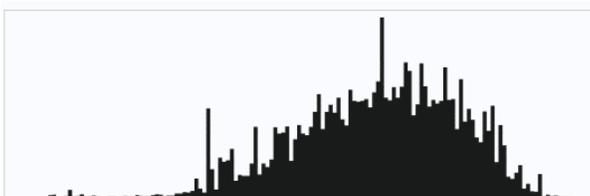

Element Quality Histogram for Extremely Coarse Grade

**Fig. S6.**

Vibration modes of spike swing in the spherical virion

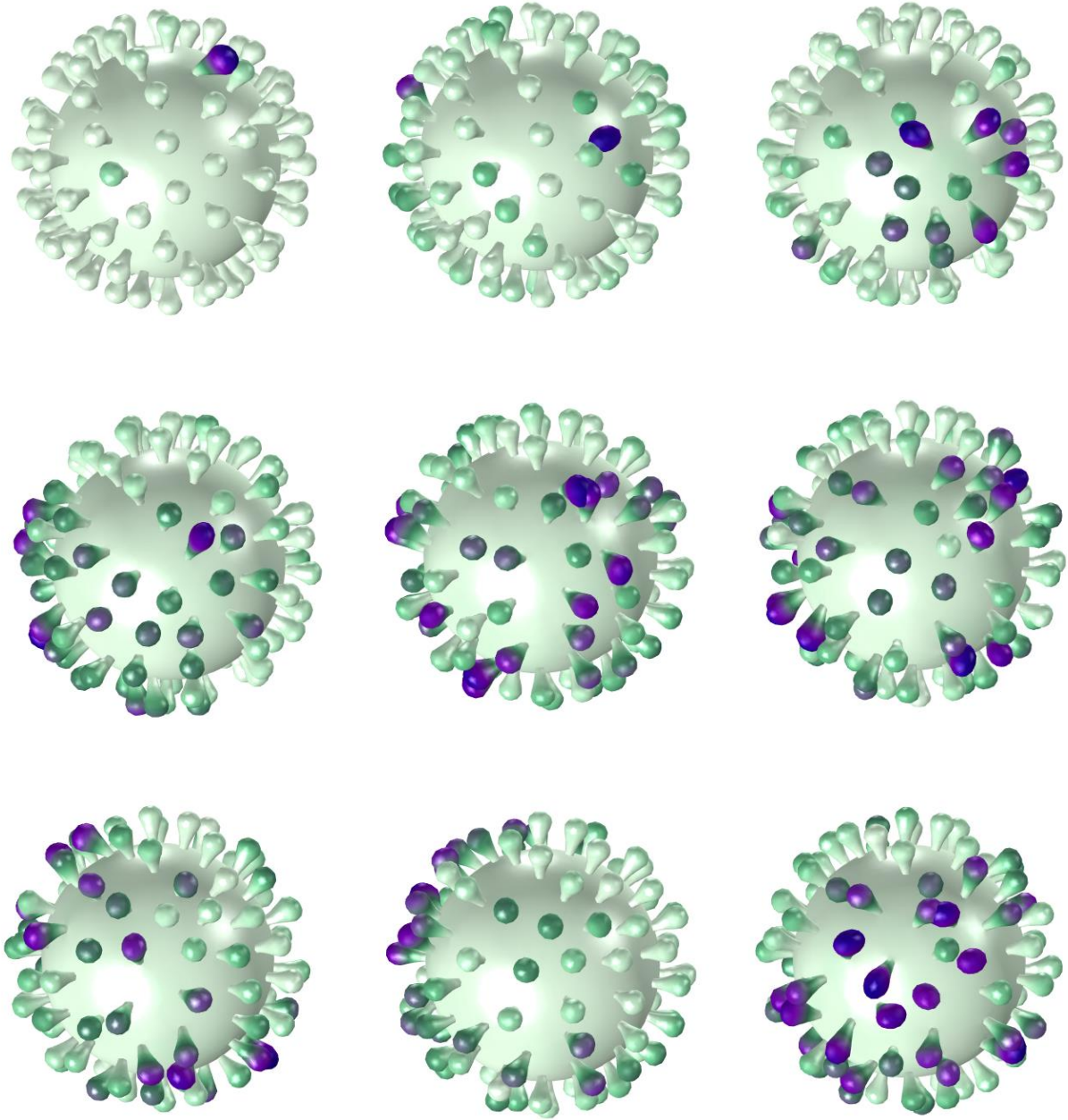

**Fig. S7.**

More vibration modes of spike swing in the spherical virion

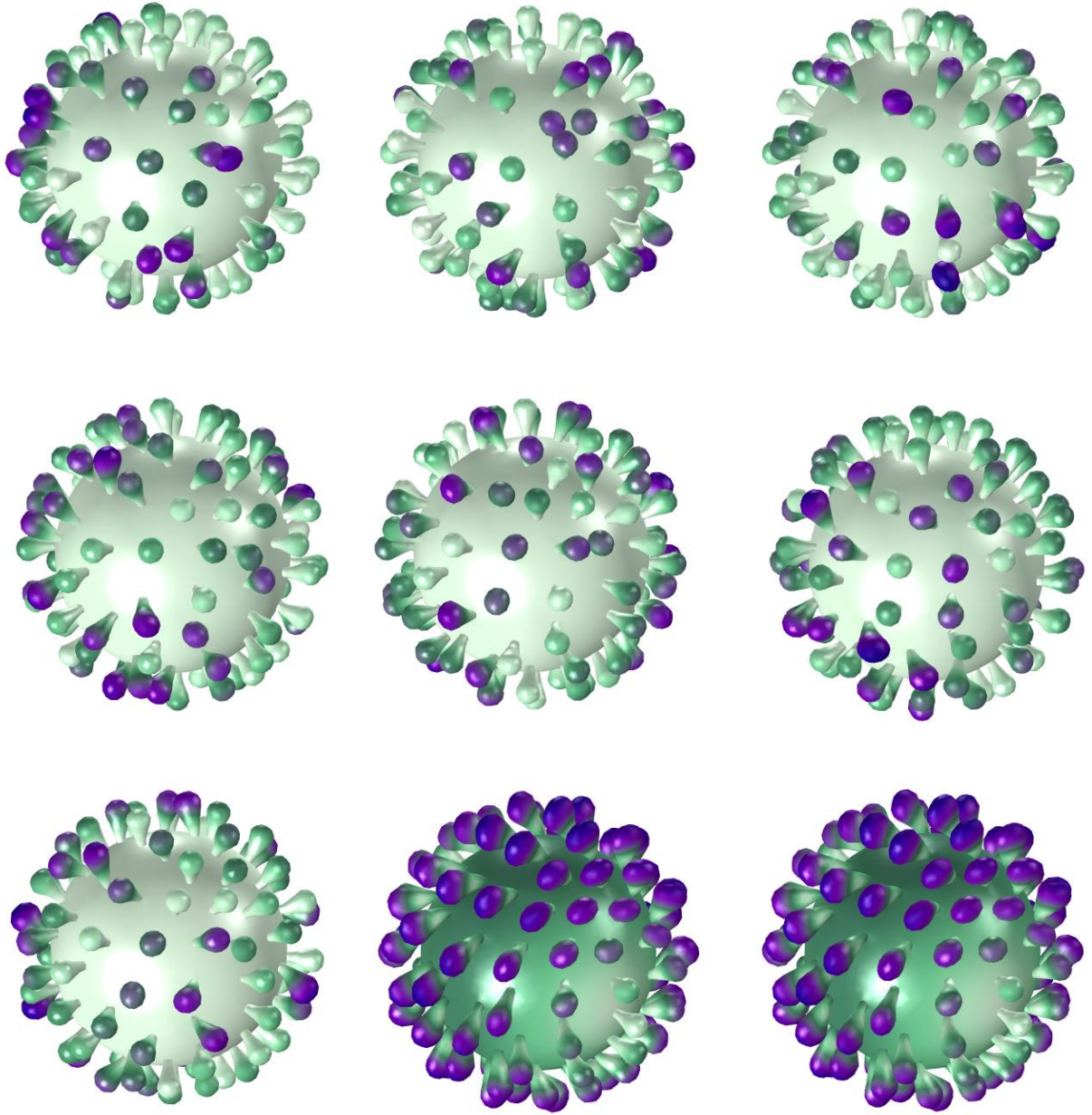

**Fig. S8.**
Vibration modes of spike swing in the ellipsoidal virion

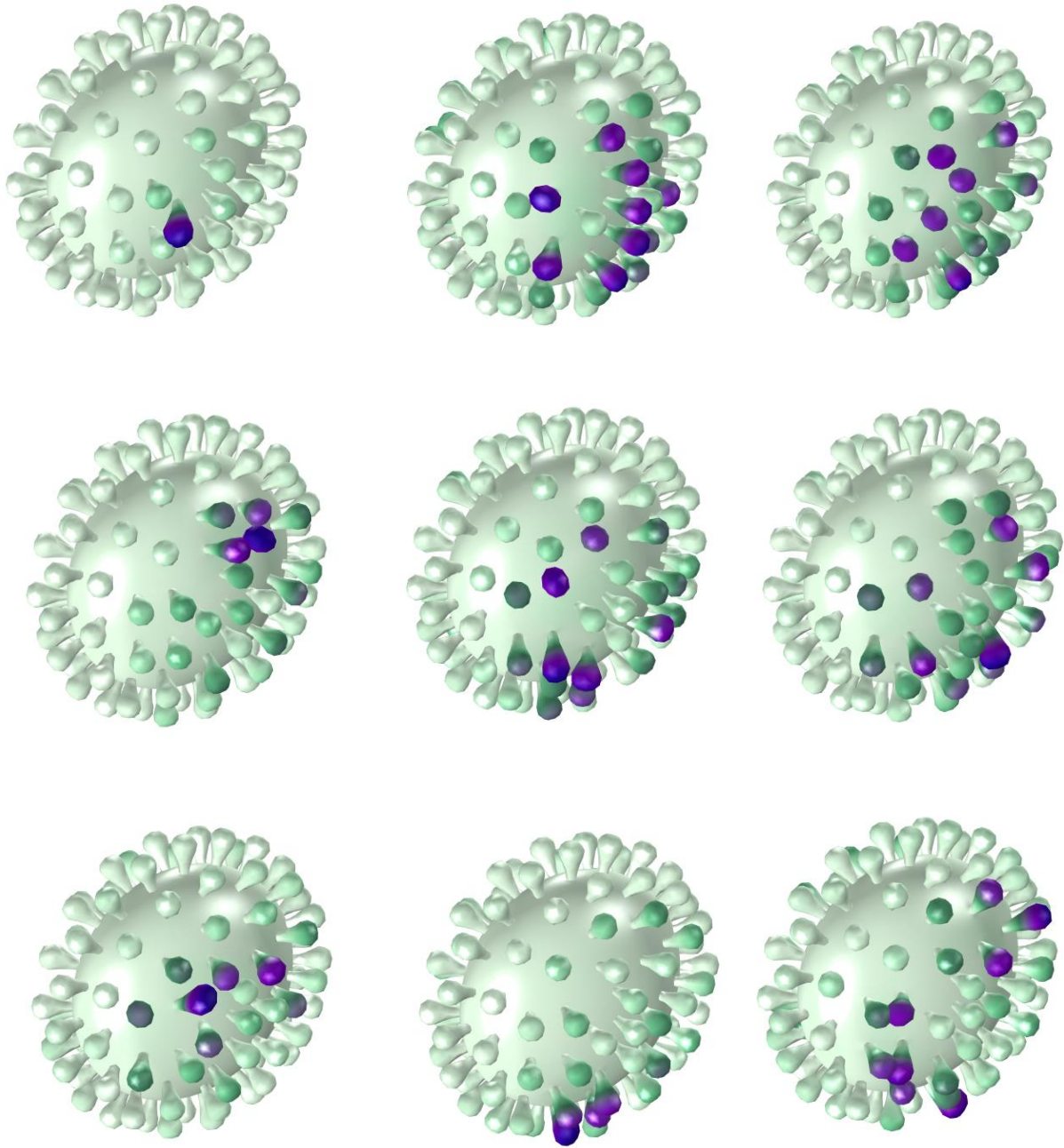

**Fig. S9.**

More vibration modes of spike swing in the ellipsoidal virion

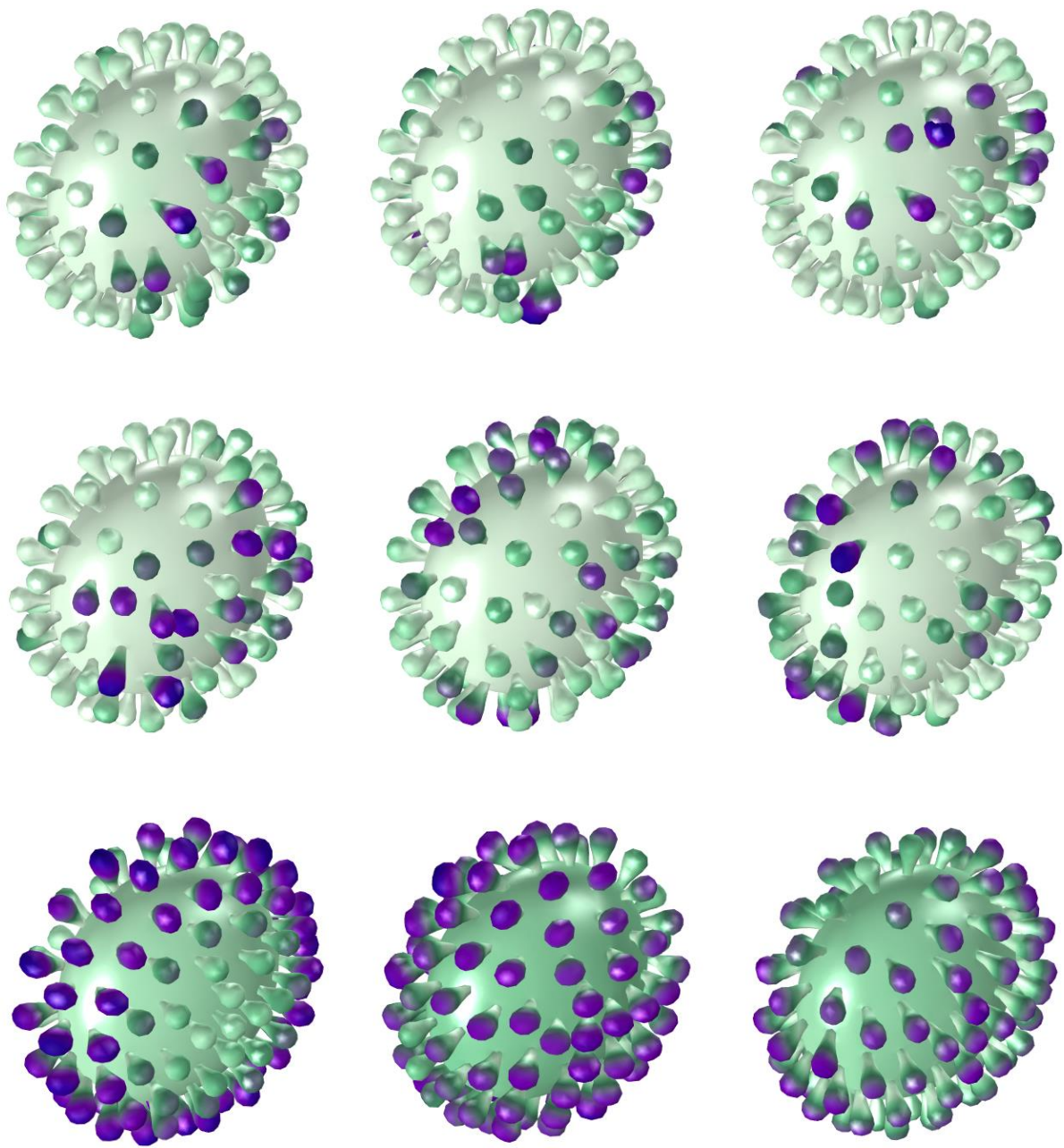

**Fig. S10.**

Stress fields stimulated by the ultrasonic wave with amplitude $P_0$=1Pa in the virion at typical instants for the case of continuously-increased $\omega_e/\omega0^{(1)}$ from 0.5 to 2.0, T=286.15K and $\zeta$=0.00

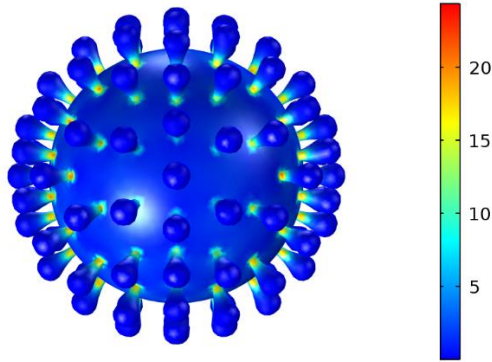

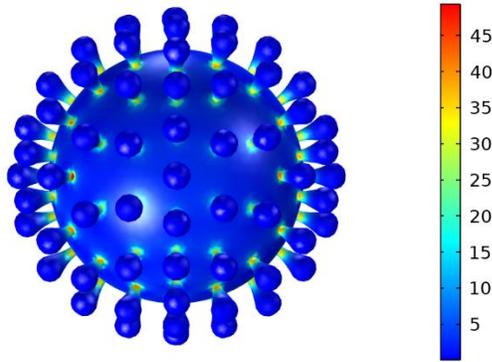

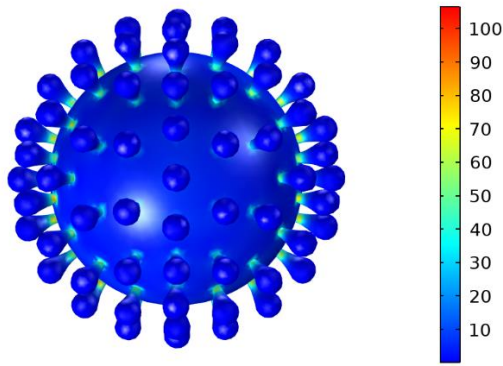

**Fig. S11.**

Stress fields stimulated by the ultrasonic wave with amplitude $P_0$=1Pa in the virion at typical instants for the case of continuously-increased $\omega_e/\omega_0^{(1)}$ from 0.5 to 2.0, T=286.15K and $\zeta$=0.01

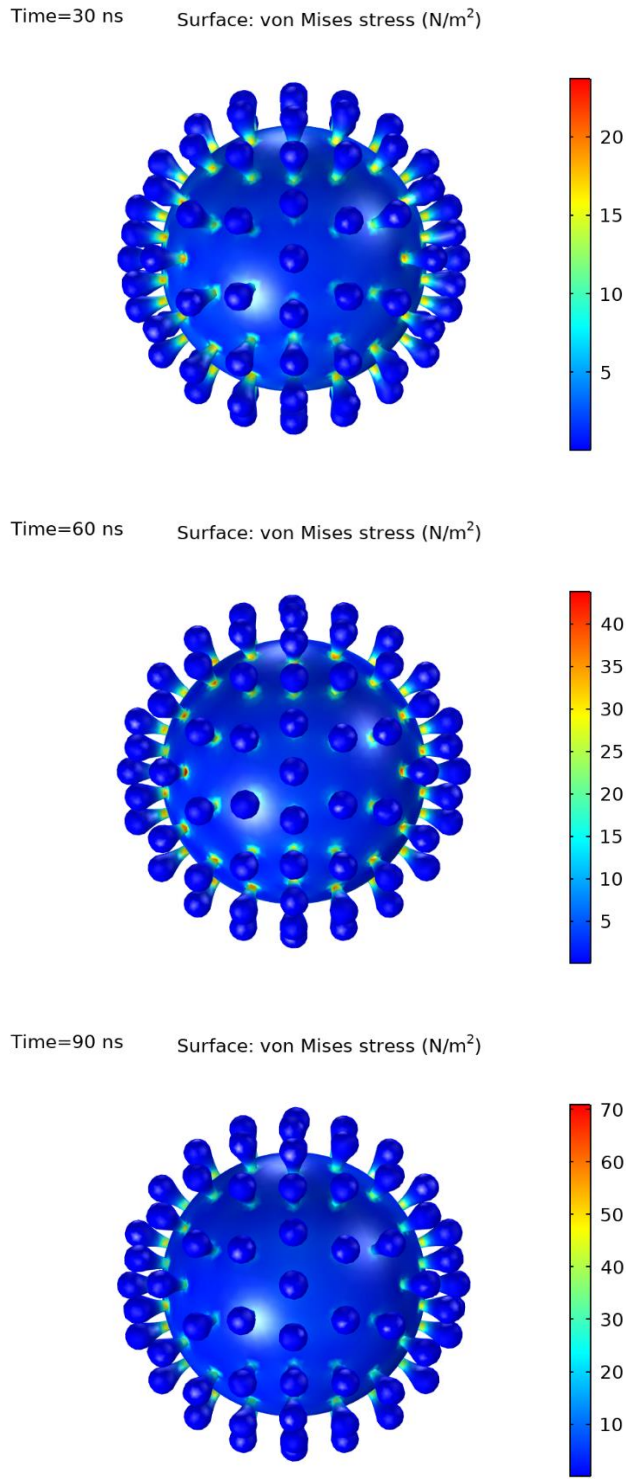

**Fig. S12.**

Stress fields stimulated by the ultrasonic wave with amplitude $P_0$=1Pa in the virion at typical instants for the case of continuously-increased $\omega_e/\omega0^{(1)}$ from 0.5 to 2.0, T=329.15K and $\zeta$=0.00

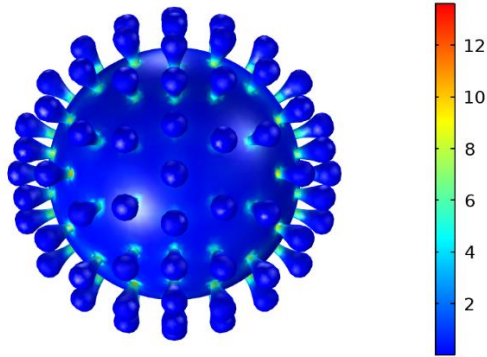

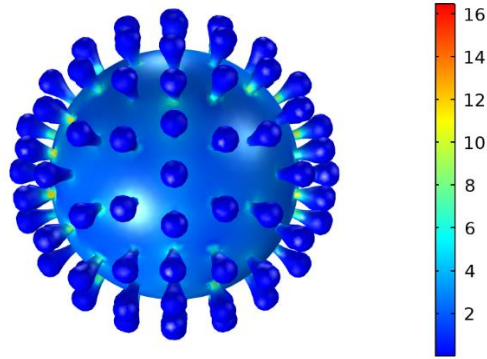

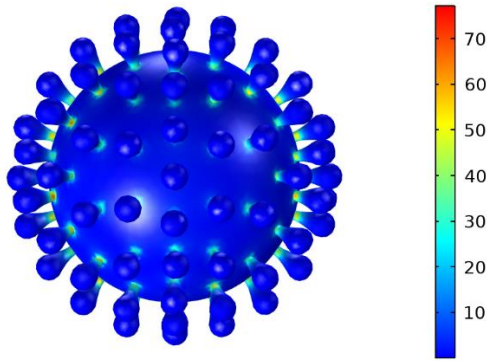

**Fig. S13.**
Stress fields stimulated by the ultrasonic wave with amplitude $P_0$=1Pa in the virion at typical instants for the case of continuously-increased $\omega_e/\omega_0^{(1)}$ from 0.5 to 2.0, T=329.15K and $\zeta$=0.01

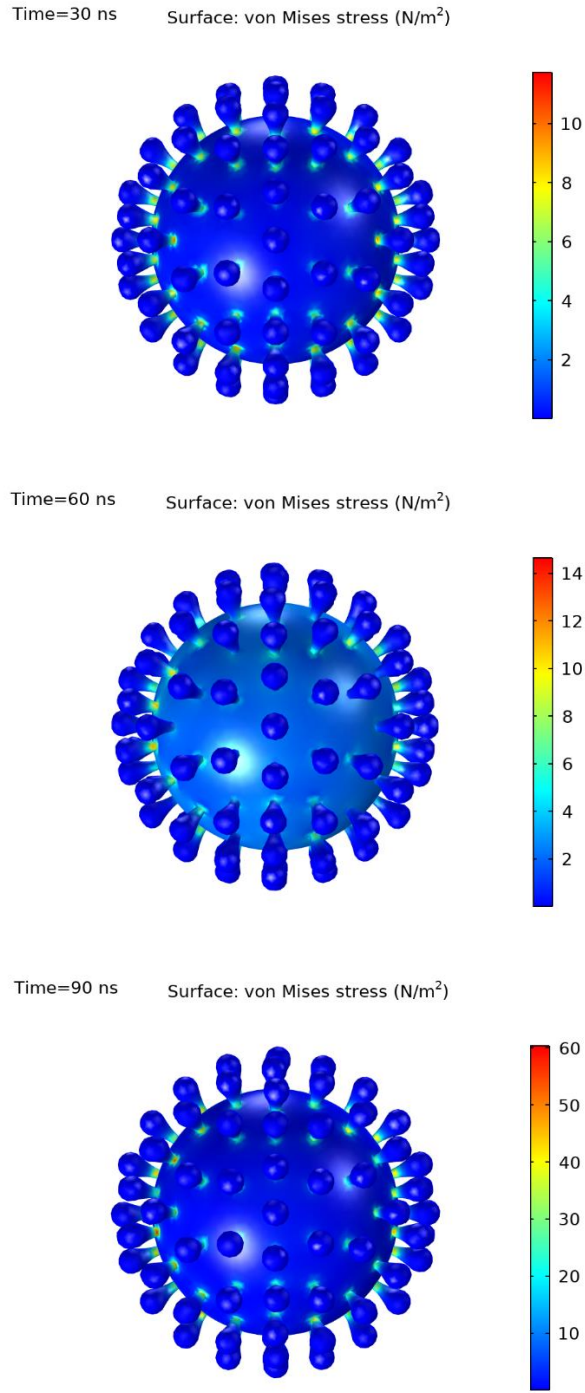

**Fig. S14.**

Stress fields stimulated by the ultrasonic wave with amplitude $P_0$=1Pa in the virion at typical instants for the case of stepwisely-increased $\omega_e/\omega0^{(1)}$ from 0.5 to 2.0, T=286.15K and $\zeta$=0.00

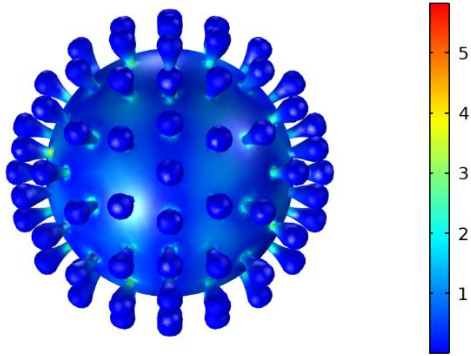

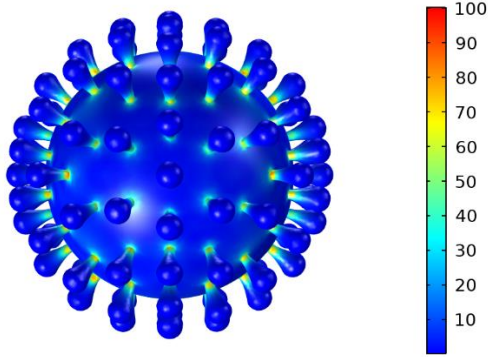

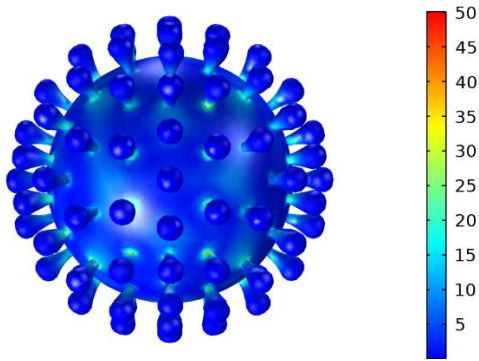

**Fig. S15.**

Stress fields stimulated by the ultrasonic wave with amplitude $P_0$=1Pa in the virion at typical instants for the case of stepwisely-increased $\omega_e/\omega_0^{(1)}$ from 0.5 to 2.0, T=286.15K and $\zeta$=0.01

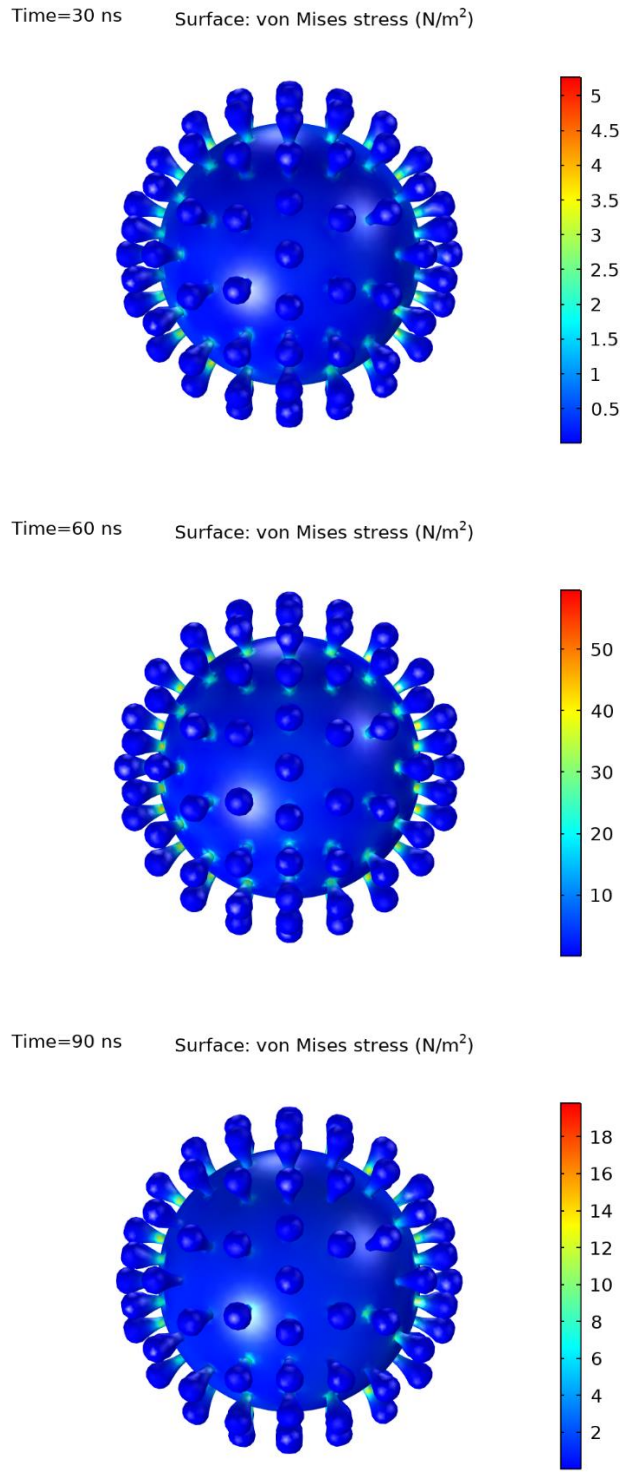

**Fig. S16.**

Stress fields stimulated by the ultrasonic wave with amplitude $P_0$=1Pa in the virion at typical instants for the case of stepwisely-increased $\omega_e/\omega0^{(1)}$ from 0.5 to 2.0, T=329.15K and $\zeta$=0.00

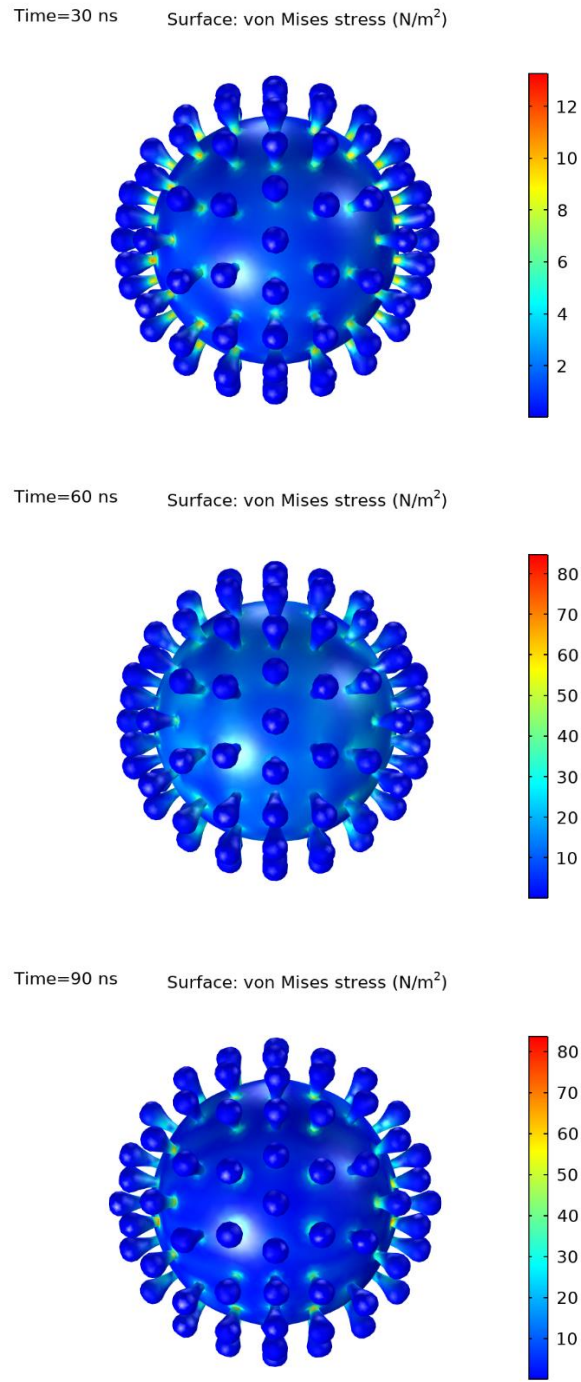

**Fig. S17.**
Stress fields stimulated by the ultrasonic wave with amplitude $P_0$=1Pa in the virion at typical instants for the case of stepwisely-increased $\omega_e/\omega0^{(1)}$ from 0.5 to 2.0, T=329.15K and $\zeta$=0.01

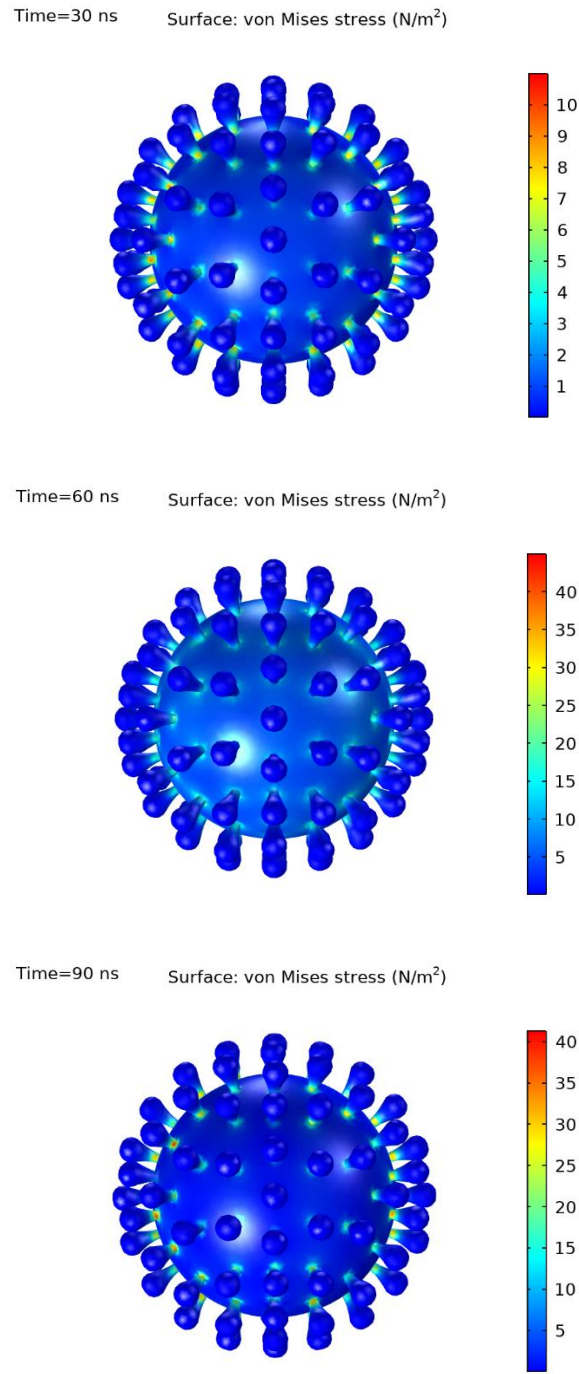

**Fig. S18.**
Maximum stress magnifications around the roots of all of the 98 spikes for the case of continuously-increased $\omega_e/\omega_0^{(1)}$ from 0.5 to 2.0, T=286.15K and $\zeta$=0.00, with time/ ns being the abscissa (horizontal axis) and the dimensionless ratio of stress magnification $\sigma_m/P_0$ ordinate (vertical axis).

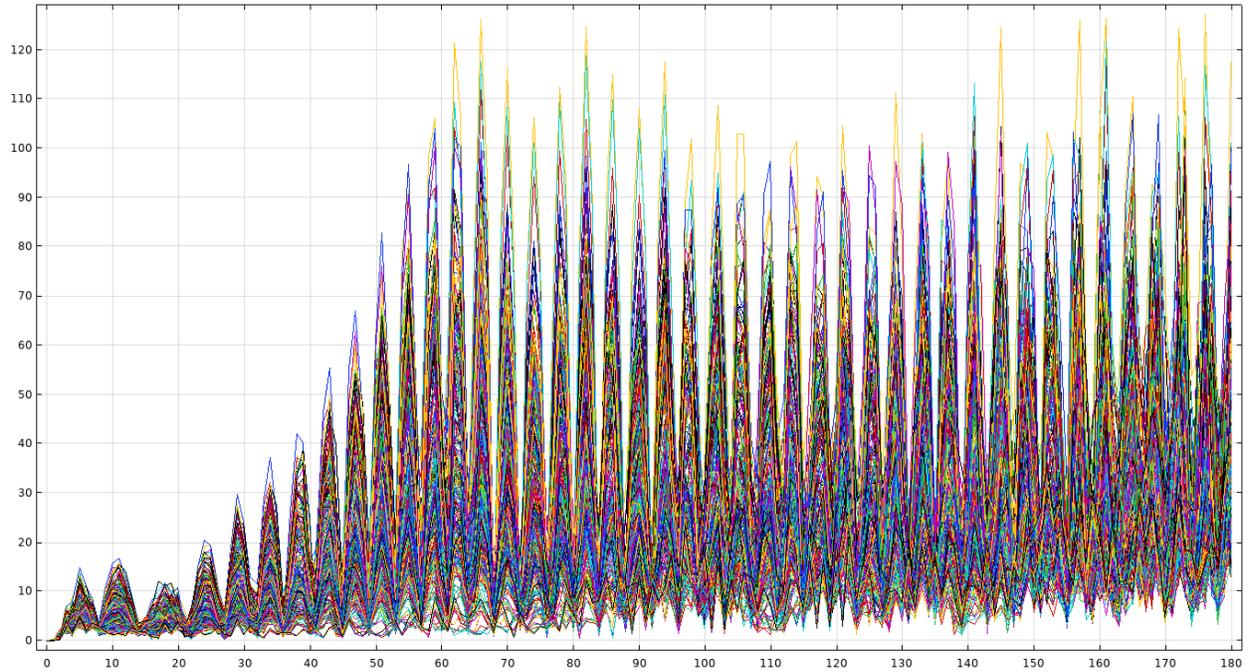

**Fig. S19.**
Maximum stress magnifications around the roots of all of the 98 spikes for the case of continuously-increased $\omega_e/\omega_0^{(1)}$ from 0.5 to 2.0, T=286.15K and $\zeta$=0.01, with time/ ns being the abscissa (horizontal axis) and the dimensionless ratio of stress magnification $\sigma_m/P_0$ ordinate (vertical axis).

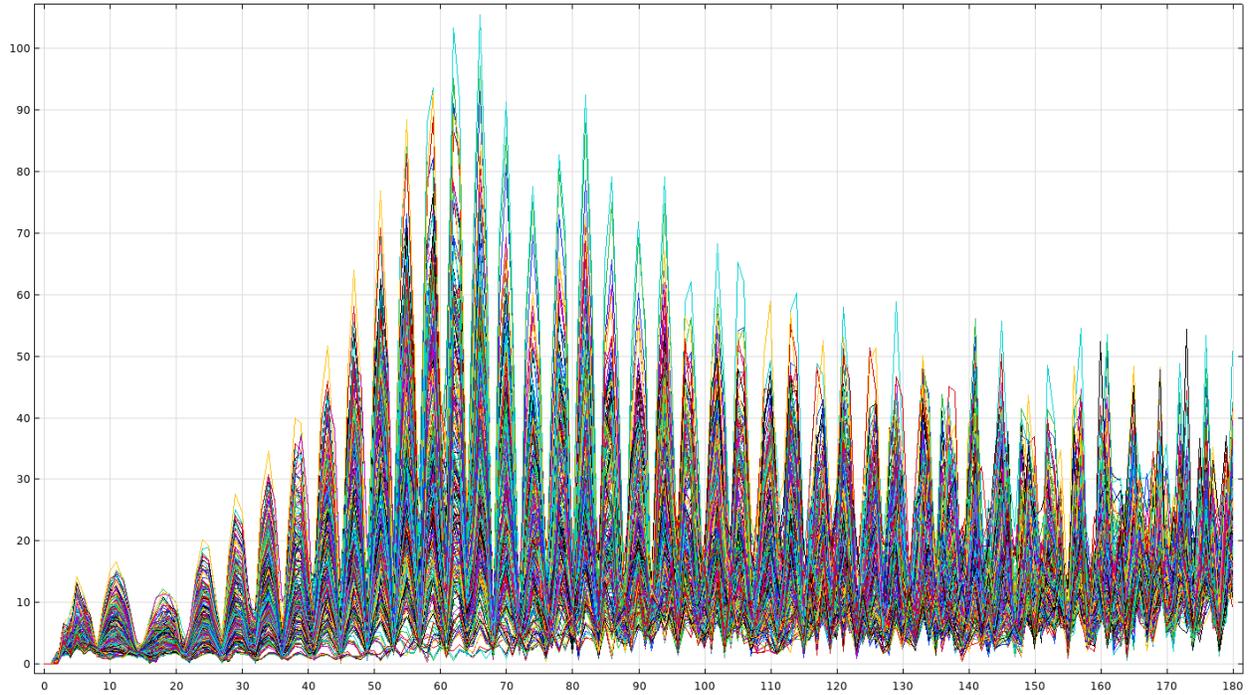

**Fig. S20.**
Maximum stress magnifications around the roots of all of the 98 spikes for the case of continuously-increased $\omega_e/\omega_0^{(1)}$ from 0.5 to 2.0, T=329.15K and $\zeta$=0.00, with time/ ns being the abscissa (horizontal axis) and the dimensionless ratio of stress magnification $\sigma_m/P_0$ ordinate (vertical axis).

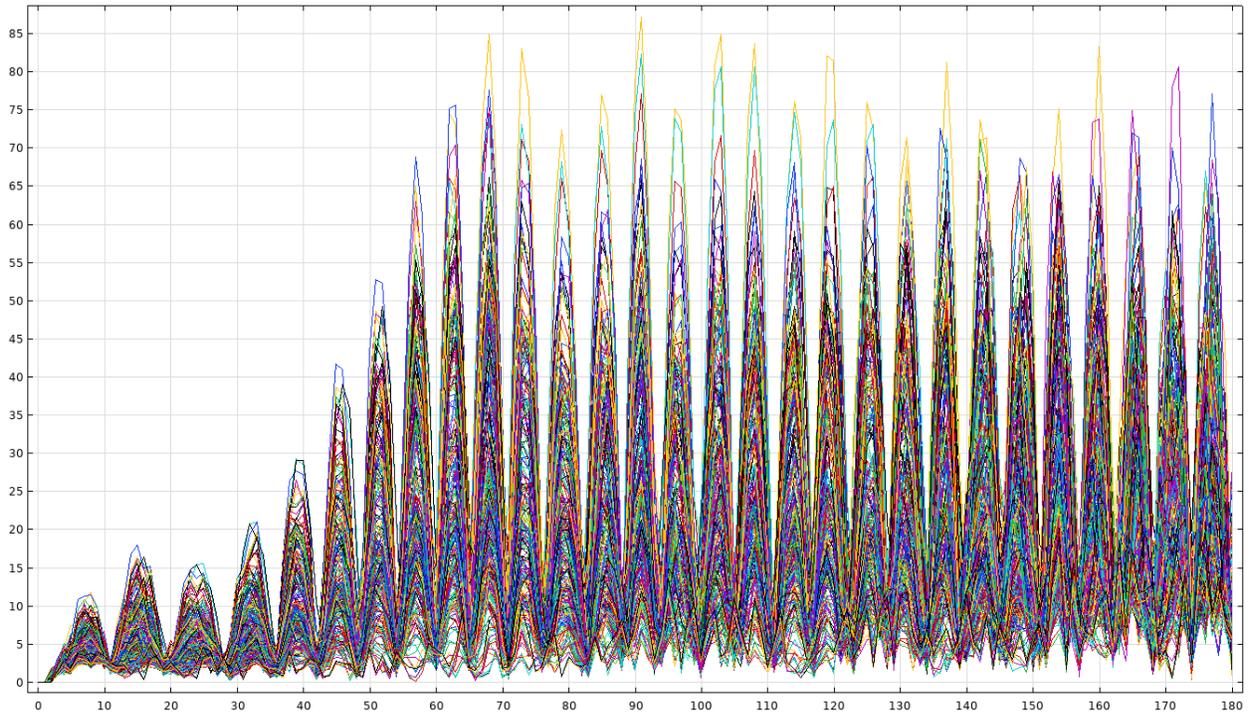

**Fig. S21.**
Maximum stress magnifications around the roots of all of the 98 spikes for the case of continuously-increased $\omega_e/\omega_0^{(1)}$ from 0.5 to 2.0, T=329.15K and ζ=0.01, with time/ ns being the abscissa (horizontal axis) and the dimensionless ratio of stress magnification $\sigma_m/P_0$ ordinate (vertical axis).

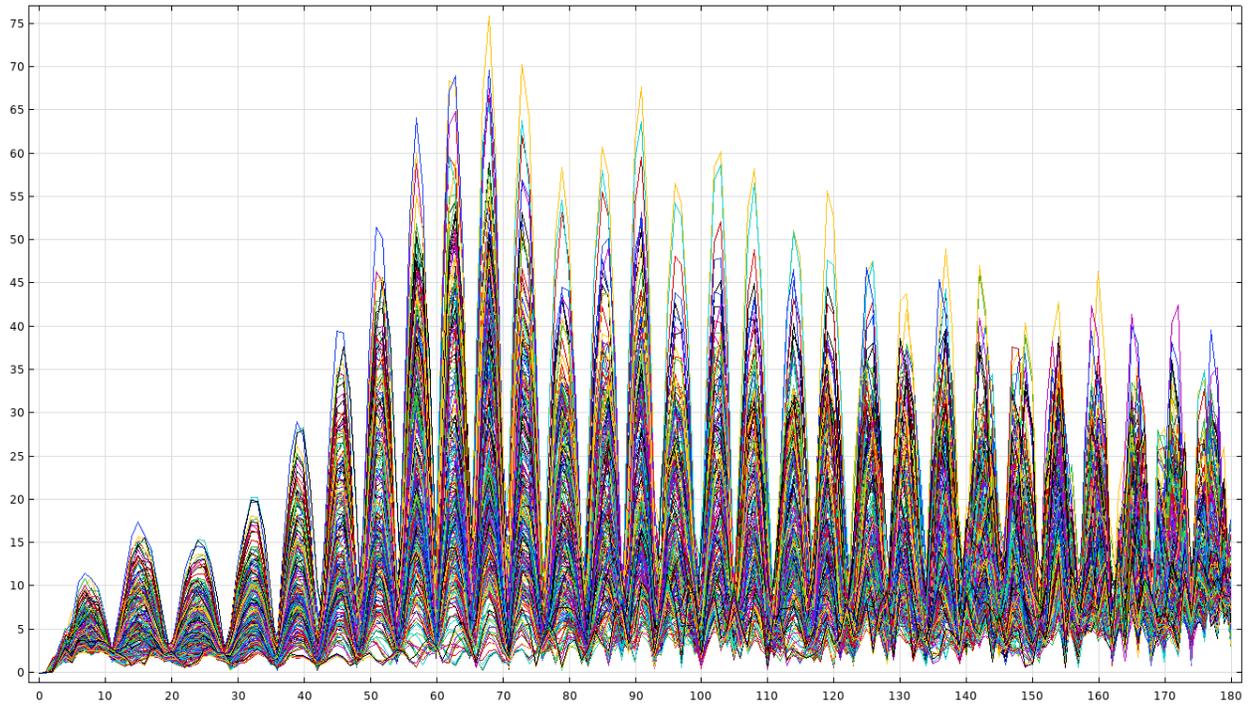

**Fig. S22.**
Maximum stress magnifications around the roots of all of the 98 spikes for the case of stepwisely-increased $\omega_e/\omega_0^{(1)}$ from 0.5 to 2.0, T=286.15K and $\zeta$=0.00, with time/ ns being the abscissa (horizontal axis) and the dimensionless ratio of stress magnification $\sigma_m/P_0$ ordinate (vertical axis).

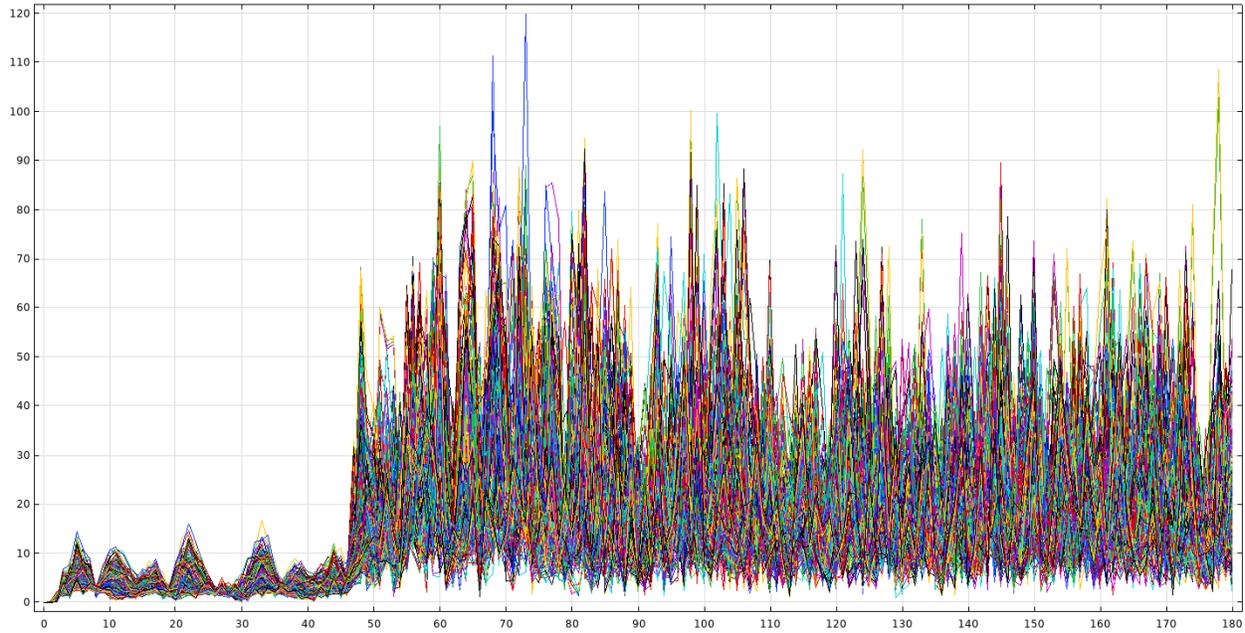

**Fig. S23.**
Maximum stress magnifications around the roots of all of the 98 spikes for the case of stepwisely-increased $\omega_e/\omega_0^{(1)}$ from 0.5 to 2.0, T=286.15K and $\zeta$=0.01, with time/ ns being the abscissa (horizontal axis) and the dimensionless ratio of stress magnification $\sigma_m/P_0$ ordinate (vertical axis).

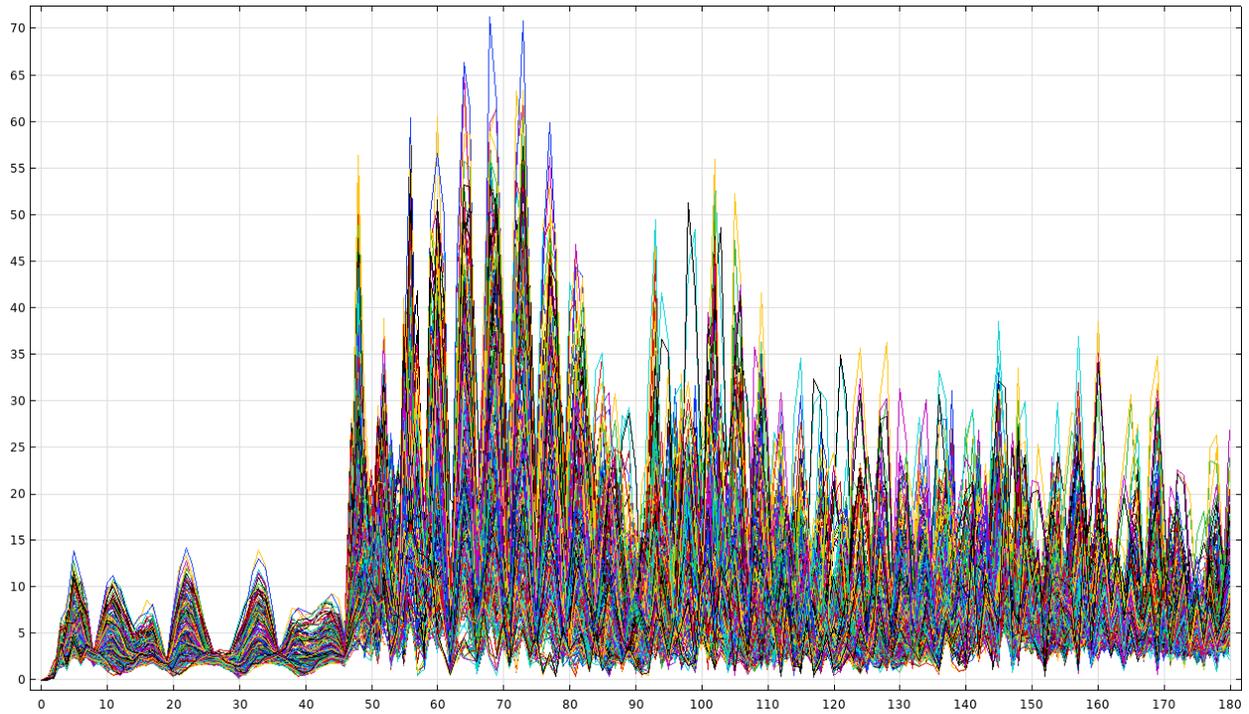

**Fig. S24.**
Maximum stress magnifications around the roots of all of the 98 spikes for the case of stepwisely-increased $\omega_e/\omega_0^{(1)}$ from 0.5 to 2.0, T=329.15K and $\zeta=0.00$, with time/ ns being the abscissa (horizontal axis) and the dimensionless ratio of stress magnification $\sigma_m/P_0$ ordinate (vertical axis).

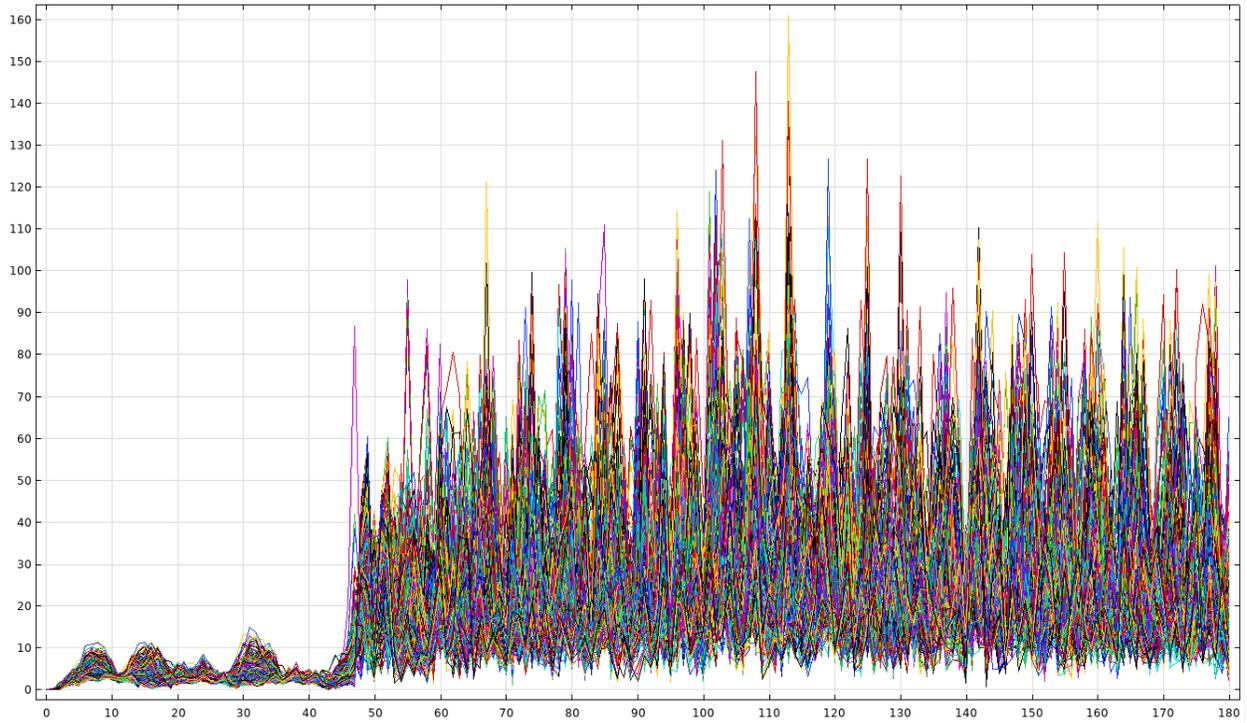

**Fig. S25.**
Maximum stress magnifications around the roots of all of the 98 spikes for the case of stepwisely-increased $\omega_e/\omega_0^{(1)}$ from 0.5 to 2.0, T=329.15K and $\zeta$=0.01, with time/ ns being the abscissa (horizontal axis) and the dimensionless ratio of stress magnification $\sigma_m/P_0$ ordinate (vertical axis).

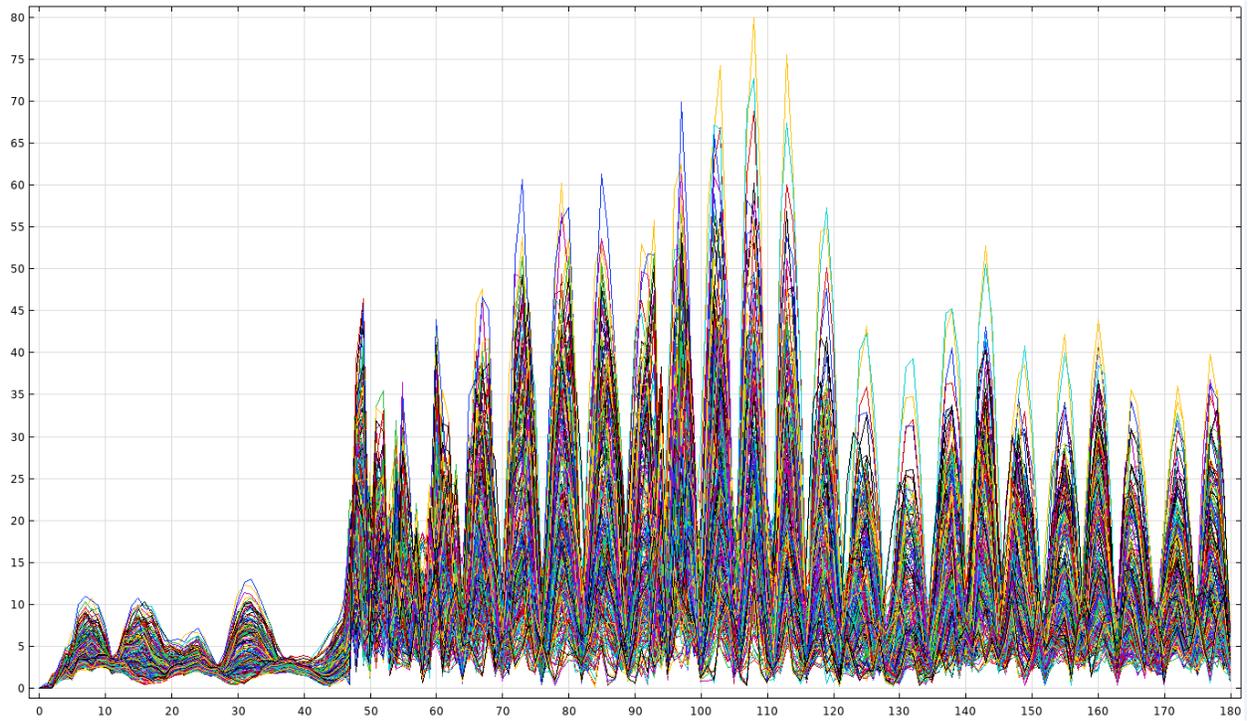

**Table S1**.
Physical parameters of the coronavirus and air *(11, 12, 17, 18, 27)*

| Virus | Density [ρ] | Poisson's ratio [ν] | Elastic modulus [E] | Damping ratio [ζ] |
|---|---|---|---|---|
| | 1378 kg/m$^3$ | 0.45 | 68MPa for temperature T=286.15K 32MPa for temperature T=329.15K | 0~0.01 |
| Air | Density [ρ] | | Speed of sound [c] | Temperature [T] |
| | 1.2 kg/m$^3$ (293.15K, 1atm) | | $\sqrt{\gamma R T}$ γ=1.4, R=287.06JKg$^{-1}$K$^{-1}$ | 286.15K~329.15K |